\documentclass{article}
\usepackage{amsmath,amsbsy,amsfonts,amssymb}
\usepackage{subfigure}
\usepackage{graphicx}
\usepackage{hyperref}

\usepackage[left=1.5cm,right=1.5cm,top=2cm,bottom=2cm]{geometry}

\begin{document}

\title{Orbital angular momentum due to modes interference}
\author{Irving Rondon $^{a\,*}$ and Francisco Soto-Eguibar $^b$  \\
$^{a}$Korea Institute for Advanced Study, School of Computational Sciences,  \\
 85 Hoegiro, Seoul 02455, Republic of Korea \\
$^{b}$  Instituto Nacional de Astrofisica Optica y Electronica, INAOE.\\
Luis Enrique Erro 1, Santa María Tonantzintla, Puebla, México 72840.\\
$^{*}$\texttt{irondon@kias.re.kr }}

\maketitle

\begin{abstract}
We present generalized expressions to calculate the orbital angular momentum for invariant beams using scalars potentials. The solutions can be separated into transversal electric TE, transversal magnetic TM and transversal electromagnetic TE/TM polarization modes. We show that the superposition of non-paraxial vectorial beams with axial symmetry can provide a well defined orbital angular momentum and that the modes superposition affects the angular momentum flux density. The results are illustrated and analyzed for Bessel beams.
\end{abstract}

\section{Introduction}
Since the pioneering and  interesting work of the orbital angular momentum presented by  L. Allen \cite{Allen}, the study  of the subject have covered  many interesting applications \cite{Alison}; a compilation overview over the last 25 years, with several theoretical and experimental applications was reported in \cite{Padgett}. The authors in \cite{Richard} have recently demonstrated that a rotationally arranged nano-antenna can be used to convert the phase information in a twisted light beam into spectral information, which hence can be used to classify the phase state of the twisted light beam. This effect has strong influence on the optical properties of dielectric and plasmonic particles and it is useful for new technological applications. The physical properties of the orbital angular momentum have the potential to improve the performance of optical communication systems in different ways \cite{Willner}, but it is not the aim of this manuscript to explore the full list of possibilities. Exploiting the physical properties of the orbital angular momentum is the subject of an increasing number of research topics nowadays. In this context, here we focus in the use of the scalar formalism for studying the orbital angular momentum for invariant beams; due to  its  simplicity, this approach has been theoretically \cite{Karem1} and experimentally \cite{Karem2,Karem3} used and tested.\\
Recently, the scalar representation has been used to describe scattering problems using partial waves series in the far field approximation, where the impinging source is a structured field \cite{Scat}. Effects  also considering the polarization \cite{Hamed}, the extinction \cite{Irving} and driven  acoustic radiation forces \cite{Rajabi} have also been considered. Furthermore, the scalar approach  is a natural representation for the structured beams family, also known as \textquotedblleft non-diffracting beams\textquotedblright. The  study of these fields covers different areas, from quantum mechanics to astrophysics \cite{Hugo}. Such fields are the Bessel  \cite{Durnin}, Mathieu  \cite{Julio} and Weber beams \cite{Bandres}, which are constructed by superposition of plane-waves \cite{Whitaker}, formalism which is called angular spectrum representation \cite{NietoV}. These ideal  fields  propagate indefinitely without changing their transverse intensity distribution \cite{Bouchal}, even in the presence of massive phase perturbations and into inhomogeneous media \cite{Florian}. As an interesting application, among others, this physical effect increases the resolution and contrast to image sub-cellular components and organelles in different microscopy methods \cite{PabloLoza}.\\
Otherwise, the study of structured invariant beams is of increasing interest in optical physics. Their properties make them particularly attractive for optical design \cite{Daniel}, for studying propagation through etalons and crystals \cite{SZhang}. Another interesting example is the use of Bessel beams for driving an optimal single tractor beam for dielectric particles with cylindrical shape \cite{Qiu}.\\
The study of nonparaxial orbital angular momentum  was recently revisited in \cite{Brandao} using Bessel beams; in that work,  the authors  show  well defined orbital angular momentum properties, without considering the possibility of   contributions due  to mixed modes  superimposed along   their  propagation.  To the best of our knowledge, this feature has not been  fully explored.  However, recently, using the scalar  potential approach, fundamentals electromagnetic properties, such as the energy density, the Poynting vector, the Maxwell stress tensor for nonparaxial beams were derived \cite{IRondon}. Currently, the study of the Poynting vector has been taken  considerable interest due to its features and  properties   along the propagation.\\
In the context of structured beams, Novitsky in \cite{Novitsky} has shown  that Bessel beams  possess negative values in the longitudinal and azimuthal components of the Poynting  vector, which depends in the superposition of mixed modes and on the phase detuning between the complex amplitudes $c_{TE}$ and $c_{TM}$ of the transversal electric or  magnetic part  of the beam representation. Since then, many interesting results have been reported, such as the explanation  of optical pulling forces \cite{Jack}, effects in meta-materials \cite{JTCosta}, tractor beams \cite{Wang} and optical manipulation \cite{Cheng}, to name a few of them. Another interesting case was reported for X-Waves in \cite{Salem}, where the propagation direction of their negative Poynting vector could be locally changed using carefully chosen complex amplitudes; however, we showed \cite{IRondon} that the negative behavior can be found independently of the mode interference. Nonetheless, the Poynting vector  behavior presented  in  the Bessel beams and  in the X-Waves, mentioned above, has opened a discussion related to tractor beams generation, and with other interesting  applications, such as the forces that  can be locally oriented in a direction opposite to the propagation wave vector \cite{Sukhov}.\\
Then,  what is the relationship between mixed modes and the Negative  Poynting Vector (NPV)? This effect can be physically defined as an uncommon response produced by the local sign change in the Poynting vector components along its propagation. Recently, we reported  in \cite{IrvingRO} to deep in the understanding of the NPV to study  the invariant family beams. The main result of our work was to show the negative local change of the negative Poynting vector, which  can be independently obtained without to superposition of  mixed modes TE/TM, as it is in the case of Weber beams \cite{IrvingRO}. Additionally, considerable interest in theoretically and experimentally studies of the vectorial structured fields are driven by the possibility to create a wide variety of exotic optical focal fields with homogeneous and spatially inhomogeneous states of polarization; an interesting review of the wide scope of interest and applications is presented in \cite{Qiwen}.\\
In this context, the study of the properties of the orbital angular momentum from the theoretical an experimental points of view can open new engineering technologies. Using the concept of the Poynting vector can be useful, as shown by  proposals to measure the orbital angular momentum using superposition of vector mixed modes TE/TM. This was performed and tested for a Laguerre Gauss beams \cite{HGarcia}. This approach  may open interest for other structured fields, as it was also proposed by using $X$ waves in ultrashort optical pulses \cite{Marco}. Even, in the case of  two electromagnetic plane waves  with the same angular frequency and  different wave vectors, the  superposition fields  reveals  highly nontrivial structure in the local momentum and spin densities \cite{Aleksandr}, that can be used to enhance the optical manipulations of small particles. For example, the spinning dynamics can be driven by superposing two vortex beams with respective circular and radial polarizations such that the particle spins around a certain optical axis \cite{Manman}. The authors in \cite{Igor,Liang}  have pointed out the importance  of the  Poynting vector  for obtaining orbital angular momentum  from spatial superposition of the Poynting  vector beams.\\
This article is organized  as follows: in Section 2, we briefly review  the theoretical framework based on the scalar potential  approach; in Section 3, we present the general negative Poynting vector for the whole invariant beams family;  in Section 4, we study the orbital angular momentum in their transversal and longitudinal propagation due to mixed modes and in Section 5, we present our conclusions.

\section{Maxwell equations in terms of scalar potentials} \label{SectionScalarPotential}
Following the formalism proposed by Stratton \cite{Stratton}, we write the electromagnetic fields as
\begin{subequations}
	\begin{align}
		\vec{E}&=c_\text{TE}\vec{M}(\vec{r}) +  c_\text{TM} \vec{N}(\vec{r}), \label{ec:VecE} \\
		\vec{H}&=-i \sqrt{\frac{\varepsilon}{\mu}}\left[ c_\text{TE}\vec{N}(\vec{r}) + c_\text{TM} \vec{M}(\vec{r})\right] \label{ec:VecH},
	\end{align}
\end{subequations}
where $\vec{M}(\vec{r})$ and $\vec{N}(\vec{r})$ are vector fields proposed as
\begin{equation}
\label{ec:VecM}
\vec{M}(\vec{r})= \nabla \times [\hat{a} \psi(\vec{r})],
\end{equation}
and
\begin{equation}
\label{ec:VecN}
\vec{N}(\vec{r})= \frac{1}{k} \nabla \times \vec{M}(\vec{r}),
\end{equation}
being $\psi$ a scalar field, $\hat{a}$ an arbitrary unit vector  that determines the direction of propagation (which we will choose as the $Z$ axis, so $\hat{a}=\hat{e}_3$),  $k$ the magnitude of the wave vector, $\varepsilon$ the electric permittivity, $\mu$ the magnetic permeability, and $c_\text{TE}$ and $c_\text{TM}$ two arbitrary complex numbers. This  approach has been successfully used to study  the properties of the family of invariant beams, theoretically and experimentally  \cite{Karem1,Karem2,Karem3}.\\
It  is straightforward to verify that if a scalar field  $\psi(\vec{r})$ satisfies the Helmholtz equation
\begin{equation}
\nabla^{2} \psi + k^{2} \psi =0,
\end{equation} 
then the fields (\ref{ec:VecE}) and (\ref{ec:VecH}) satisfy the Maxwell equations; so, the scalar field $\psi(\vec{r})$ will be  named scalar potential. Note that the vector fields, $\vec{M}$ and $\vec{N}$, are orthogonal, that is $\vec{M} \cdot \vec{N}=0$, and solenoidal, i.e. $\nabla \cdot \vec{M}=0$ and $\nabla \cdot \vec{N}=0 $.\\
On the other hand, for any invariant beam the spatial evolution of  the scalar potential $\psi$  can be described by its transverse and longitudinal parts \cite{Boyer}. The transverse part $\varphi(u_1, u_2)$ will depend only on the transverse coordinates, $u_1,u_2$, and the longitudinal part $Z(z)$ will depend on the longitudinal coordinate $z$ (as we selected $\hat{a}=\hat{e}_3$), physically the propagation axis; i.e., we can write
\begin{equation}
\label{ec:VarSepa}
\psi(u_1,u_2,z)=\varphi(u_1,u_2)Z(z).
\end{equation}
After  substituting (\ref{ec:VarSepa}) in the Helmholtz equation, we easily obtain that $\varphi(u_1, u_2)$ satisfies the two dimensional transverse Helmholtz equation
\begin{equation} \label{ec:HemholtzTrans}
\nabla _T^2\varphi + k^2_T \varphi=0,
\end{equation}
where $\nabla _T^2$ is the Laplacian transversal operator, which has a specific form in each coordinate system, and the longitudinal part is $Z(z)=e^{i k_z z}$ with the dispersion relation $k^2= k_T^2 + k_z^2$. The two dimensional transverse Helmholtz equation \eqref{ec:HemholtzTrans} can be separated in Cartesian, cylindrical, parabolic cylindrical and elliptical cylindrical coordinates \cite{Boyer}, and that gives origin to plane waves, Bessel beams, Weber beams and Mathieu beams, respectively.  Then, we can write the vector operator fields, Eqs. \eqref{ec:VecM} and \eqref{ec:VecN}, as follows 
\begin{equation}
\label{ec:OpM}
\vec{M}=-e^{i k_z z}\nabla _T^{\bot }\varphi,
\end{equation}
where
\begin{equation}\label{nablatp}
\nabla _T^{\bot }=-\hat{e}_1\frac{1}{h_2}\frac{\partial }{\partial u_2}+\hat{e}_2\frac{1}{h_1}\frac{\partial }{\partial u_1},
\end{equation}
$ \hat{e}_1 $ and $\hat{e}_2$ are the base unit vectors corresponding to the transversal direction, and $h_1$ and $h_2$ are the corresponding scale factors.  We note that in the four coordinate systems, in which equation \eqref{ec:HemholtzTrans} can be separated, the scale factor $ h_3$ is equal to 1. It is also easy to verify that
\begin{equation}
\label{ec:OpN}
\vec{N}= \frac{e^{i k_z z}}{k}\left(  i k_z \nabla _T+\hat{e}_3 k_T^2\right)\varphi,
\end{equation}
where $\nabla _T$ is the transversal part of the $\nabla$ operators, i.e. 
\begin{equation}
\label{nablat}
\nabla _T=\hat{e}_1\frac{1}{h_1}\frac{\partial }{\partial u_1}+\hat{e}_2\frac{1}{h_2}\frac{\partial }{\partial u_2}.
\end{equation}
It is worth to notice that the transversal vector operators are related as follows $\nabla_{T}^{\perp}=\hat{e}_3\times\nabla_{T}$.

\section{ A generalized Poynting vector for scalars potentials}
The Poynting vector represents the directional power flux per unit area of an electromagnetic field. For harmonic electromagnetic fields the time average of the Poynting vector is given by \cite{Jackson}
\begin{equation} \label{ec:VecPoyn}
\left\langle  \vec{S}  \right\rangle= \frac{1}{2} \mathrm{Re} \left(\vec{E} \times \vec{H}^* \right).
\end{equation}
Substituting expressions \eqref{ec:VecE} and \eqref{ec:VecH} for the electromagnetic fields into \eqref{ec:VecPoyn}, a generalized expression for the Poynting vector of any  invariant beam is obtained \cite{IRondon},
\begin{align}\label{GenVecPoynt}
	\left\langle  \vec{S} \right\rangle = \left|  c_{\text{TE}}\right| ^2\left\langle  \vec{S}_{\text{TE}} \right\rangle 
	+\left|  c_{\text{TM}}\right| ^2 \left\langle  \vec{S}_{\text{TM}} \right\rangle 
	+\left\langle \vec{S}_{\text{TE/TM}}\right\rangle ,
\end{align}
where
\begin{equation}\label{STE}
\left\langle \vec{S}_{\text{TE}}\right\rangle = \frac{1}{2 k }\sqrt{\frac{\varepsilon}{\mu}} \mathbf{Re}
\left[\left(\nabla_T\varphi \cdot \nabla_T\varphi^\ast\right)k_z \hat{e}_3  -i   k_T^2 \, \varphi^* \, \nabla _T\varphi\right] 
\end{equation}
is the transversal electric part,
\begin{equation}\label{STM}
\left\langle \vec{S}_{\text{TM}}\right\rangle = \frac{1}{2 k }\sqrt{\frac{\varepsilon}{\mu}} \mathbf{Re}
\left[\left(\nabla_T\varphi \cdot \nabla_T\varphi^\ast\right)k_z \hat{e}_3 +i   k_T^2 \, \varphi^* \, \nabla _T\varphi\right] 
\end{equation}
is the transversal magnetic part, and the interference modes TE/TM
\begin{align}\label{Sint}
\left\langle \vec{S}_{\text{TE/TM}}\right\rangle  = &
\frac{1}{2k^2} \sqrt{\frac{\varepsilon}{\mu}}  \mathbf{Re} \Big[  i 
\left( c_{\text{TE}}\,c_{\text{TM}}^* k^2+   c_{\text{TE}}^*\,c_{\text{TM}}  k^2_z \right)   \left(     \nabla_T \varphi^*\times \nabla_T \varphi  \right)    \cdot  \hat{e}_3 
+c_{\text{TE}}^*\,c_{\text{TM}}      k_z k_T^2  \,  \hat{e}_3 \times \nabla_T   \left(  \varphi^* \varphi   \right) \Big]. 
\end{align}
The time averaged Poynting vector   is independent of the $z$ coordinate and it satisfies  $\nabla\cdot \left\langle  \vec{S}  \right\rangle =0$, as was proven in \cite{Horak}. It is important to remark that the interference  term in  the Poynting vector, expressed in Eq. \eqref{Sint}, is different from zero for any invariant beam, of course, whenever at least one of the two constants $c_{\text{TE}}$ and $c_{\text{TM}}$ is not null. Physically this confirms the negative behavior trough the propagation; these results  were recently reported in \cite{IrvingRO}  using the Weber beams. The study of its effect on the orbital angular momentum is done in the following sections.

\section{Orbital angular momentum density}
For harmonic electromagnetic fields,  the time averaged linear momentum  per unit volume carried is $\left\langle \vec{p}\right\rangle =\left\langle \vec{S}\right\rangle/c^2$ \cite{Jackson,Peter}, and then, the time averaged orbital angular momentum density is $\left\langle \vec{j}\right\rangle =\vec{r} \times \left\langle \vec{p}\right\rangle=\vec{r} \times \left\langle \vec{S}\right\rangle/c^2$; using Eqs. \eqref{STE}, \eqref{STM} and \eqref{Sint}, we can find explicitly the contributions from the transverse electric, from the transverse magnetic and from the transverse electric/transverse magnetic TE/TM mixed modes. The transversal electric TE part is
\begin{align}	\label{Jte}
\left\langle \vec{j}^{\,\,\text{TE}}  \right\rangle  
= \frac{\varepsilon}{2 \omega}   \left|  c_{\text{TE}}\right| ^2  \mathbf{Re} 
\left[ k_z \left(   \nabla _{T }  \varphi \cdot \nabla _{T} \varphi^* \right)  \vec{r} \times\hat{e}_3  -i   k_{T}^2 \varphi^* \, \vec{r} \times \nabla _{T}\varphi\right],
\end{align}
whereas  the transversal magnetic TM  contribution has the form
\begin{align}\label{Jtm}
\left\langle \vec{j}^{\,\,\text{TM}}  \right\rangle 
= \frac{\varepsilon}{2 \omega}   \left|  c_{\text{TM}}\right| ^2  \mathbf{Re} 
\left[ k_z \left(   \nabla _{T }  \varphi \cdot \nabla _{T} \varphi^* \right)  \vec{r} \times\hat{e}_3  + i   k_{T}^2 \varphi \, \vec{r} \times \nabla _{T}\varphi^*\right],
\end{align}
and the interference mixed modes TE/TM part is given by
\begin{align}\label{Jtetm1}
\left\langle \vec{j}^{\; \; \text{TE/TM}}  \right\rangle= \frac{c \varepsilon }{2 \omega ^2}   \mathbf{Re} \Biggl\{ i \left( k^2 c_{\text{TE}} c^*_{\text{TM}}+ c^*_{\text{TE}} c_{\text{TM}} k_z^2 \right)\left(     \nabla_T \varphi^*\times \nabla_T \varphi  \right)    \cdot (\vec{r} \times  \hat{e}_3) 
+c_{\text{TE}}^*\,c_{\text{TM}} k_z k_{T}^2   \,
\vec{r}\times [\hat{e}_3 \times \nabla_T   \left(  \varphi^* \varphi   \right)]  \Biggr\}.
\end{align}
The set of equations  \eqref{Jte}, \eqref{Jtm} and \eqref{Jtetm1} provide  a useful simple recipe to calculate the orbital angular momentum of any invariant beam in terms of scalar potentials. It is important to remark that Eq. \eqref{Jtetm1} physically represents the orbital angular momentum propagation due to the superposition of modes TE/TM.\\

\subsection{Example: orbital angular momentum for Bessel beams using  a scalar potential}
The properties of  the orbital angular momentum  have been extensively investigated by different means \cite{Padgett,GMendez}. However, to the best of our knowledge, the study of the orbital angular momentum arising from mode superposition  has not been attempted before; nevertheless, the interference Bessel beams have been applied in the micro-manipulation of particles \cite{McGloin,Mazilu}. With this in mind, and as an illustrative example, we analyze the particular case of the orbital angular momentum of Bessel beams.\\
The transversal solution of the scalar Helmholtz equation in cylindrical coordinates is\\
\begin{equation}\label{BB}
\varphi(r,\theta)=  J_{m} \left( k_{T}  r\right)   e^{i m \theta},
\end{equation}
where $m$ is any integer, $J_{m}(\zeta)$ is the Bessel function of the first kind of order $m$, and $k_T=k\sin\beta$ is the transversal vector \cite{Hugo}. Substituting \eqref{BB} in \eqref{Jte}, \eqref{Jtm} and \eqref{Jtetm1}, we obtain the orbital angular momentum of a Bessel beam. The radial angular momentum is 
\begin{align}\label{jr}
\left\langle j_{r}  \right\rangle=\frac{c \varepsilon  }{ 2 \omega ^2}
\frac{  zk_{T}^2 }{ r }
\Biggl\{  -\left(\left| c_{\text{TE}}\right| ^2+\left| c_{\text{TM}}\right|^2\right)   k m J^2_{m }\left(rk_{T}\right)
+ 2 k_z \mathbf{Re} \left( c^*_{\text{TE}} c_{\text{TM}}\right) J_{m}\left(rk_{T}\right) \left[ rk_{T} J_{m -1}\left(rk_{T}\right)- m  J_{m }\left(rk_{T}\right) \right] \Biggr\},
\end{align} 
while the azimuthal angular momentum component is
\begin{align}\label{jtheta}
\left\langle j_{\theta}  \right\rangle
=& \frac{\varepsilon}{2 r \omega }  
\Biggl\{ - \left( \left| c_{\text{TE}}\right|^2+\left| c_{\text{TM}}\right|^2\right) k_z \left[ r^2k_{T}^2 J^2_{m -1}\left(rk_{T}\right)+2 m ^2 J^2_{m}\left(rk_{T}\right)-2 m  rk_{T} J_{m -1}\left(rk_{T}\right) J_{m}\left(rk_{T}\right) \right] 
\\ \nonumber&
+ 2 k m  J_{m}\left(rk_{T}\right) \left[ rk_{T} J_{m -1}\left(rk_{T}\right)- m  J_{m}\left(rk_{T}\right) \right] \mathbf{Re} \left( c_{\text{TE}} c^*_{\text{TM}}+ c^*_{\text{TE}} c_{\text{TM}}\frac{ k_z^2}{k^2}\right)\Biggr\},
\end{align}
and the longitudinal component is
\begin{align}\label{jz}
\left\langle j_z  \right\rangle = \frac{c \varepsilon  }{2 \omega ^2}k_{T}^2
\Biggl\{ \left(\left| c_{\text{TE}}\right|^2+\left| c_{\text{TM}}\right|^2\right) m \, k   J^2_{m }\left(rk_{T}\right) 
+2 k_z \mathbf{Re} \left(c^*_{\text{TE}} c_{\text{TM}}\right) J_{m } \left(rk_{T}\right)
\left[ rk_{T} J_{m -1} \ \left(rk_{T}\right)-m  J_{m}\left(rk_{T}\right) \right] \Biggr\}.
\end{align}
As can be observed, all terms have a clear contribution of modes superposition, which in general is different from zero. Even, for the most simple case, when the azimuthal value is zero,  $m=0$, the mixed modes are given by 
\begin{equation}
\left\langle j_{r}  \right\rangle =   \mathbf{Re} \left(c^*_{\text{TE}} c_{\text{TM}}\right) \frac{c \varepsilon}{\omega ^2}   k_z  k_T^3  z  J_0\left(r k_T\right) J_1\left(r k_T\right),
\end{equation}
\begin{equation}
\left\langle j_{\theta } \right\rangle= \left( \left|c_{\text{TE}}\right|^2+\left| c_{\text{TM}}\right|^2\right)\frac{\varepsilon     }{2 \omega } k_z k_T^2  r J_1^2\left(r k_T\right),
\end{equation}
\begin{equation}
\left\langle j_z \right\rangle = - \mathbf{Re} \left(c^*_{\text{TE}} c_{\text{TM}}\right) \frac{c \varepsilon   }{\omega ^2}  k_z  k_T^3  r J_0\left(r k_T\right)  J_1\left(r k_T\right).
\end{equation}
If the mixing is considered, these results show that there is interference between the radial and azimuthal components. Writing $c_\text{TE}= \vert c_\text{TE} \vert e^{i\phi_1} $ and $c_{\text{TM}}= \vert c_\text{TM} \vert e^{i\phi_2} $, we have $\mathbf{Re}[ c_\text{TE}^* c_{\text{TM}}]= \cos{ (\phi_1-\phi_2)}$; thus this term contribute if there is a difference of phase between $c_\text{TE}$ and $c_\text{TM}$, otherwise this term is zero. Therefore, all the well-know results for the orbital angular momentum can be obtained.

\subsection{Longitudinal orbital angular momentum}
\begin{figure}
	\centering
	\includegraphics[width=0.45\textwidth]{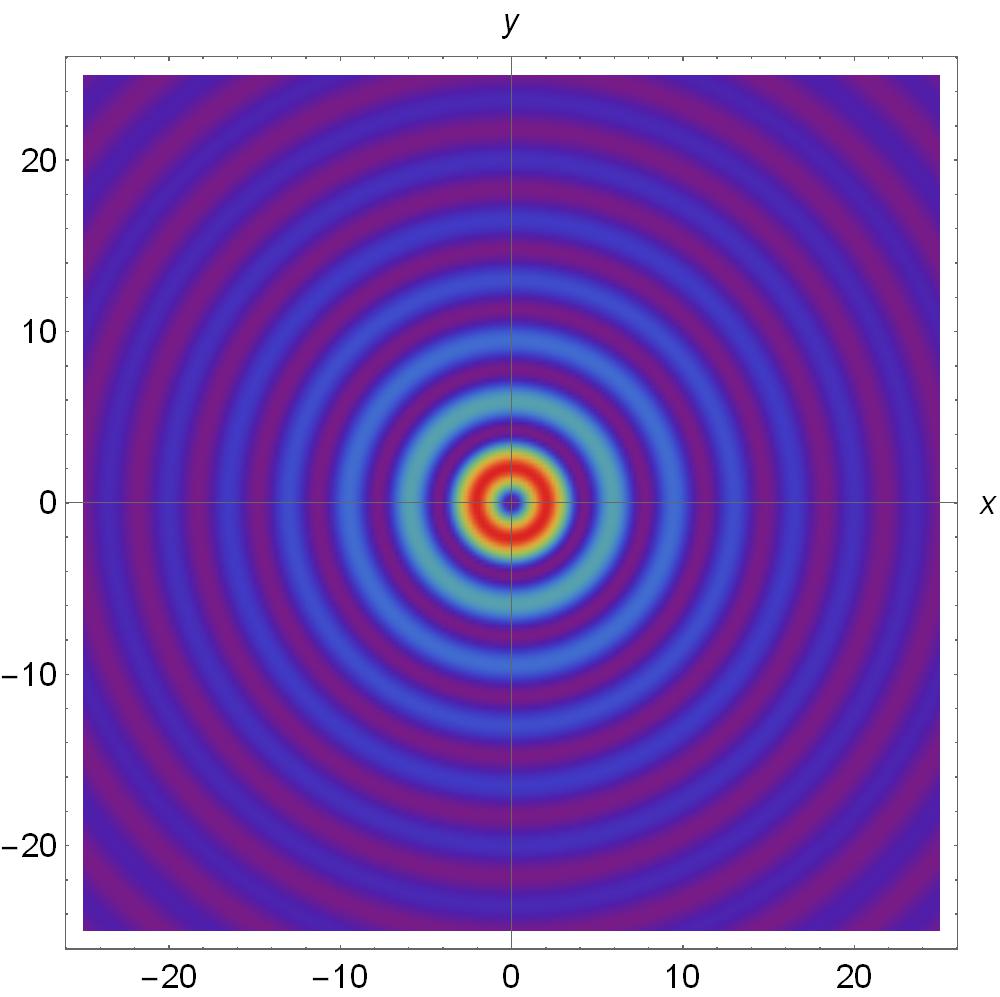}
	\qquad
	\includegraphics[width=0.45\textwidth]{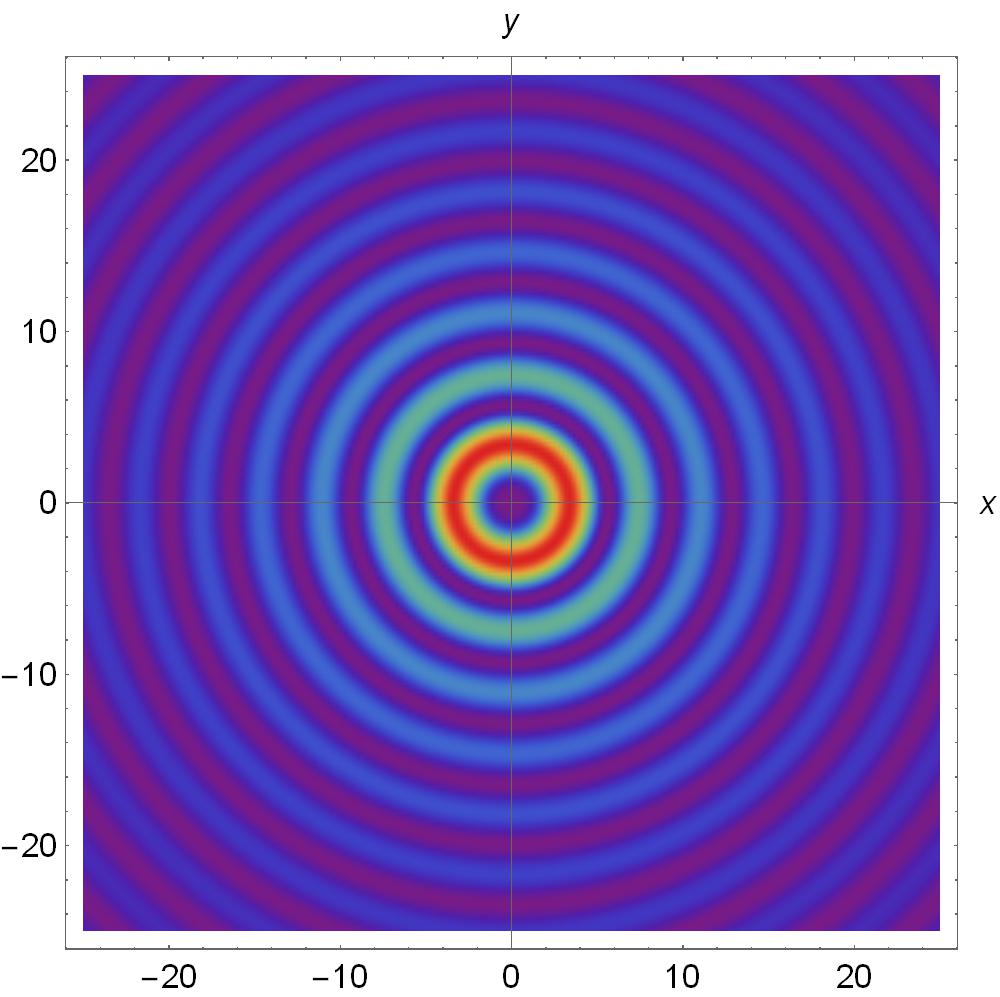}
	\caption{Longitudinal orbital angular momentum $\vert \left\langle j_{z}^{\,\,\text{TE}}  \right\rangle \vert$ of a transversal electric  Bessel beam with $m=1$ (left) and of a transversal magnetic $ \vert \left\langle j_{z}^{\,\,\text{TM}}  \right\rangle \vert$ with $m=2$ (right). } \label{Figure1}
\end{figure}

\begin{figure}
	\centering
	\includegraphics[width=0.45\textwidth]{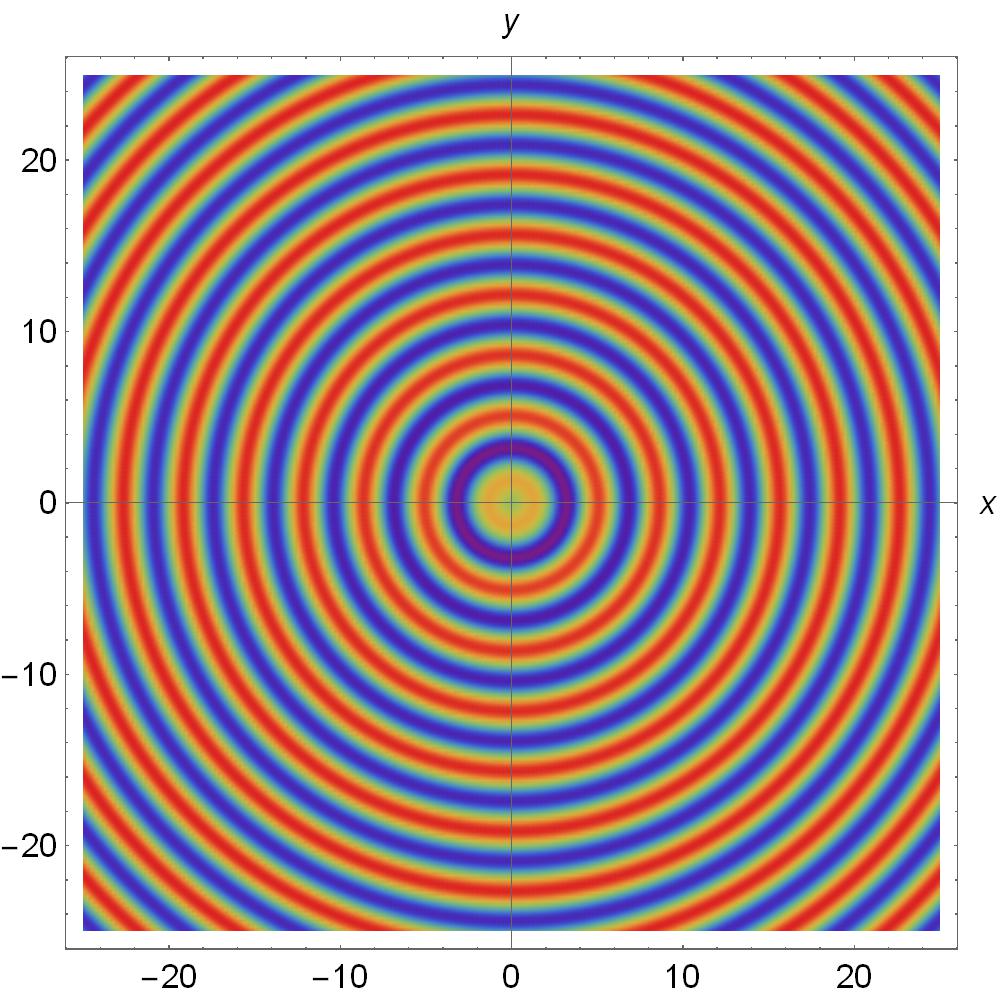}
	\qquad
	\includegraphics[width=0.45\textwidth]{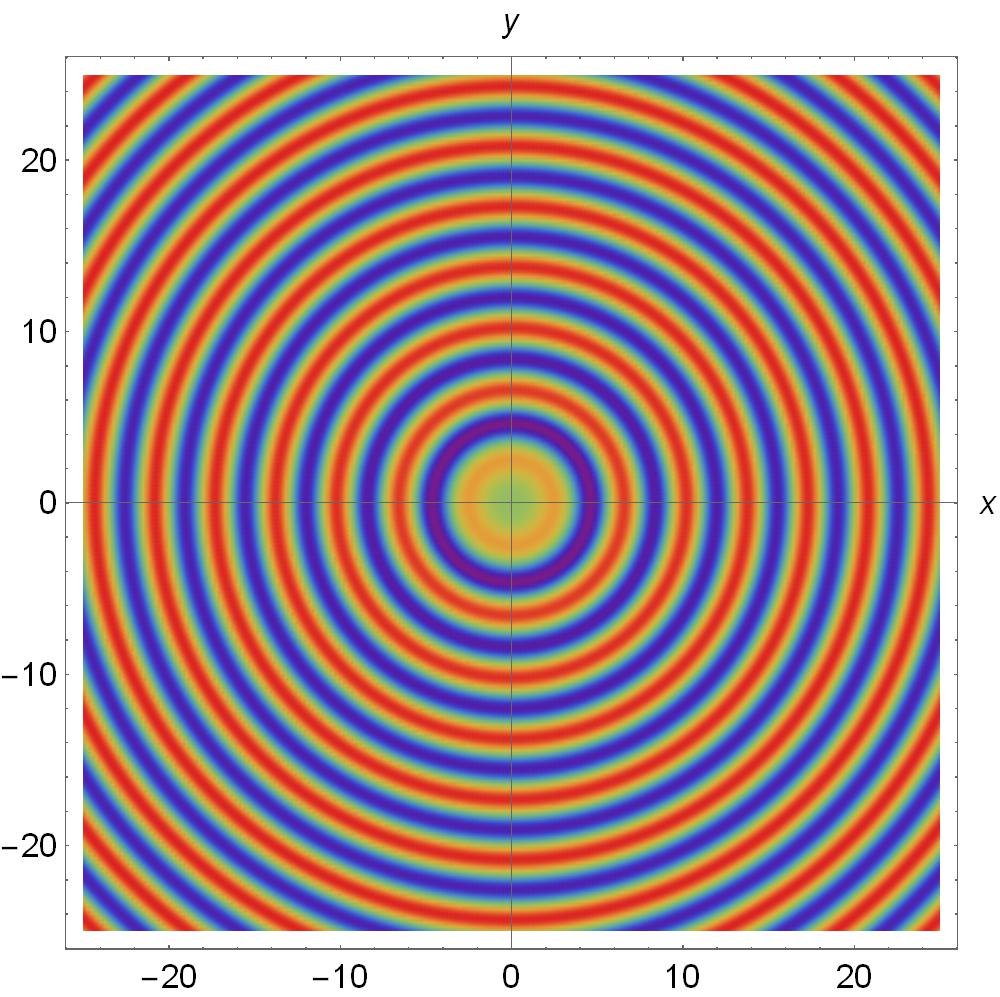}
	\caption{ The longitudinal orbital angular momentum interference term $\vert \left\langle j_{z}^{\,\,\text{TE/TM}}  \right\rangle \vert $ of a  Bessel beam, the left one with $m=1$ and the right one with $m=2$.} 	\label{Figure2}
\end{figure}

The above equations allow us to obtain the  longitudinal orbital angular momentum  for a particular transversal electric, magnetic or interference mode; the procedure is just to take the scalar product  of  \eqref{Jte}, \eqref{Jtm} and \eqref{Jtetm} with $\hat{e}_3$. For any invariant beams the longitudinal orbital angular momentum can be obtained as  
\begin{align}\label{jzTE}
\left\langle j_{z}^{\,\,\text{TE}}  \right\rangle=  
\frac{ \varepsilon k_{T}^2}{2 \omega}
\left|  c_{\text{TE}}\right|^2  \mathbf{Re}\left\{ \left( \varphi^* \vec{\mathcal{L}}_{T} \varphi \right) \cdot\hat{e}_{3} \right\} ,   
\end{align}
\begin{align}\label{jzTM}
\left\langle j_{z}^{\,\,\text{TM}}  \right\rangle= - 
\frac{ \varepsilon k_{T}^2}{2 \omega}
\left|  c_{\text{TM}}\right|^2    \mathbf{Re} \left\{ \left( \varphi \vec{\mathcal{L}}_{T} \varphi^*\right)\cdot\hat{e}_{3}  \right\},
\end{align}
\begin{align}\label{jzTETM}
\left\langle j_{z}^{\,\,\text{TE/TM}}  \right\rangle =  \frac{ \varepsilon }{2 \omega}
\frac{k_z}{k} k_{T}^2 \mathbf{Re}
\left\{ c_{\text{TE}}^*\,c_{\text{TM}} \left( \vec{\mathcal{L}}_{T}^{\perp} \vert \varphi \vert^2 \right)\cdot\hat{e}_{3} \right\},
\end{align}
where  $\vec{\mathcal{L}}_{T} = -i \vec{r}\times \nabla_{T}$  and $\vec{\mathcal{L}}_{T}^{\perp} =  \vec{r}\times \nabla_{T}^{\perp}$. Where $\nabla_{T}^{\perp}$ is given by Eq. \eqref{nablatp} and which physically means a $\pi/2$ rotation of $\nabla_T$. Note that only the transversal structure beam, which is given for  a single scalar potential $\varphi$, is required; all terms are proportional to $k_T^2$. The equation  \eqref{jzTETM} is given  in terms of mixed modes and it is proportional to the ratio $k_z/k$. In the particular case of the Bessel beams, we substitute expression \eqref{BB} into  \eqref{jzTE} and \eqref{jzTM} obtaining the following results
\begin{align}\label{jzTEm}
\left\langle j_{z}^{\,\,\text{TE}}  \right\rangle=  
\frac{ \varepsilon k_{T}^2}{2 \omega}
\left|  c_{\text{TE}}\right|^2 m  J_m^2\left(r k_T\right) ,   
\end{align}
and
\begin{align}\label{jzTMm}
\left\langle j_{z}^{\,\,\text{TM}}  \right\rangle= 
\frac{ \varepsilon k_{T}^2}{2 \omega}
\left|  c_{\text{TM}}\right|^2  m  J_m^2\left(r k_T\right) .
\end{align}
We have obtained a well-known  defined orbital angular momentum;  these results recover the results reported in \cite{Brandao} for Bessel beams. A well defined longitudinal orbital angular momentum is obtained for TE and TM  modes when $m=0$, since there is no orbital angular momentum  in the direction of propagation; if $m \neq 0$, the beam carries orbital angular momentum in the direction of  propagation \cite{Alison,Padgett}. Additionally, using the Eq. \eqref{jzTEm}, Eq. \eqref{jzTMm} and the longitudinal electromagnetic energy density reported in \cite{IRondon}, it is straightforward to  verify that the ratio of   $\left\langle j_{z}^{\,\,\text{TE}}  \right\rangle / \left\langle U \right\rangle_{z}=\left\langle j_{z}^{\,\,\text{TM}}  \right\rangle / \left\langle U \right\rangle_{z} \propto m/\omega$ \cite{Brandao}. These results confirm a  well-defined values of angular momentum  and energy for this nonparaxial approximation.\\ For the case of longitudinal mixed modes, substituting \eqref{BB} into  \eqref{jzTETM} and after some  algebraic manipulation, we get
\begin{align}\label{jzTETMm}
\left\langle j_{z}^{\,\,\text{TE/TM}}  \right\rangle = \frac{ \varepsilon }{ \omega}
\frac{k_z}{k} k_{T}^2 \mathbf{Re}
[ c_{\text{TE}}^*\,c_{\text{TM}} ] \Bigg(  \frac{d }{dr}[r J_m^2 (k_T r)] - J_m^2 (k_T r) \Bigg).
\end{align}
In this more general case, the longitudinal orbital angular momentum is not proportional  to the topological charge $m$, as in the single mode case.
Now it is related  to the radial derivative of the square field  weighted by its radius minus the  intensity of the incident field.\\
In Fig. \ref{Figure1}, it is shown the  transverse electric  $\vert \left\langle j_{z}^{\,\,\text{TE}}  \right\rangle \vert $ and  the transverse magnetic $\vert \left\langle j_{z}^{\,\,\text{TM}}  \right\rangle \vert$  longitudinal orbital angular momentum for  Bessel beams with $m=1$ and $m=2$. 
\noindent In Fig. \ref{Figure2} the mixed modes superposition $\left\langle j_{z}^{\,\,\text{TE/TM}}  \right\rangle$ longitudinal orbital angular momentum is shown for $m=1$ and for $m=2$. We obtain a  well-defined regions due to the interference superposition modes can be observed. This feature can be useful; the application of Bessel beams for optical manipulation has been proposed \cite{McGloin}, and it  has recently been  successfully developed for the case of a single-beam or counter-propagating beam trapping \cite{Chenglong}.
\noindent Recently, in \cite{AAiello}, it has been shown how combining two  Bessel beams, as in  equation \eqref{BB}, with topological charges  $m = \pm 1$, it is possible  to  generate a Hermite-Gauss (HG) beam, which posses well defined  orbital angular momentum \cite{Kotlyar}. Moreover, it is important to mention that a superposition of Hermite-Gauss beams can be transformed into a Laguerre-Gauss beam \cite{Yanfeng} which posses a very well-know and characteristic $\exp({im\phi})$ factor \cite{Allen,Alison,Padgett}. Thus, our generalized analytical formulation makes possible the study the orbital angular momentum using a single  scalar  potential.

\subsection{The transversal  interference term carry orbital angular momentum }
\begin{figure}
	\centering
	\includegraphics[width=0.45\textwidth]{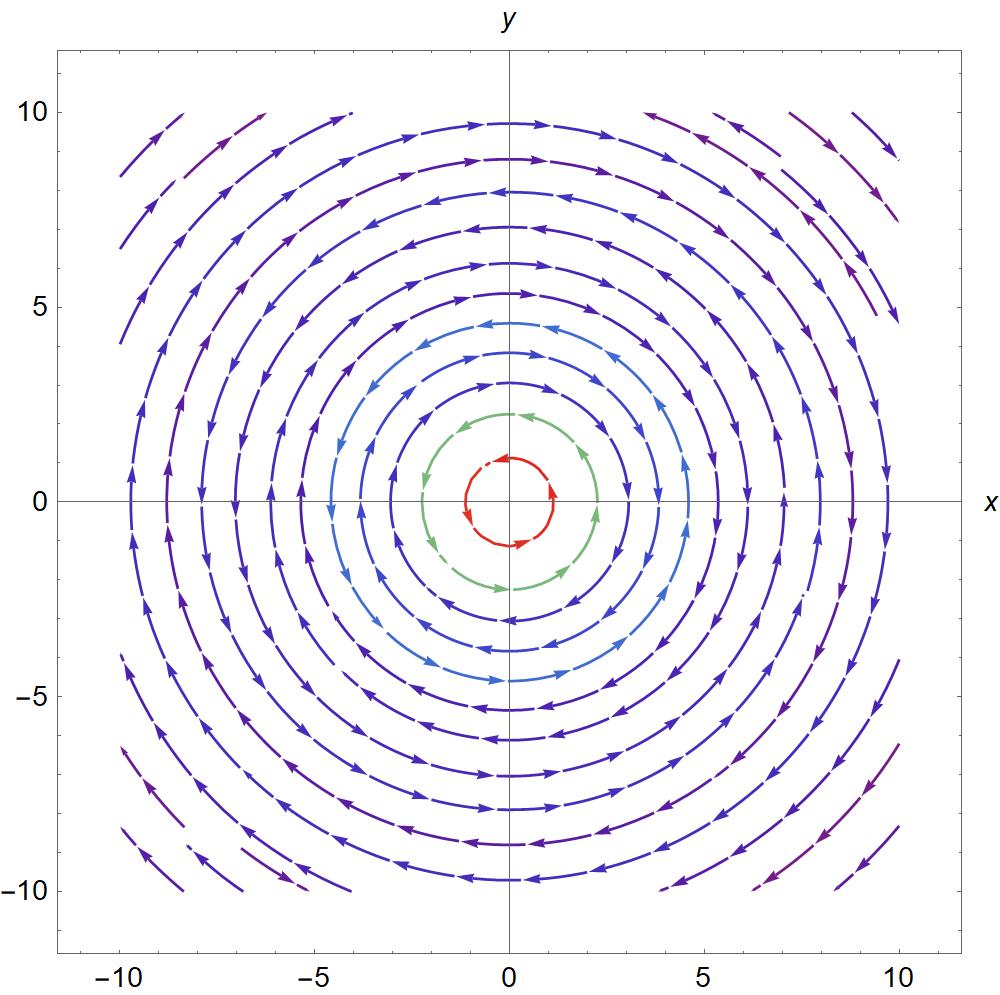} \qquad 
	\includegraphics[width=0.45\textwidth]{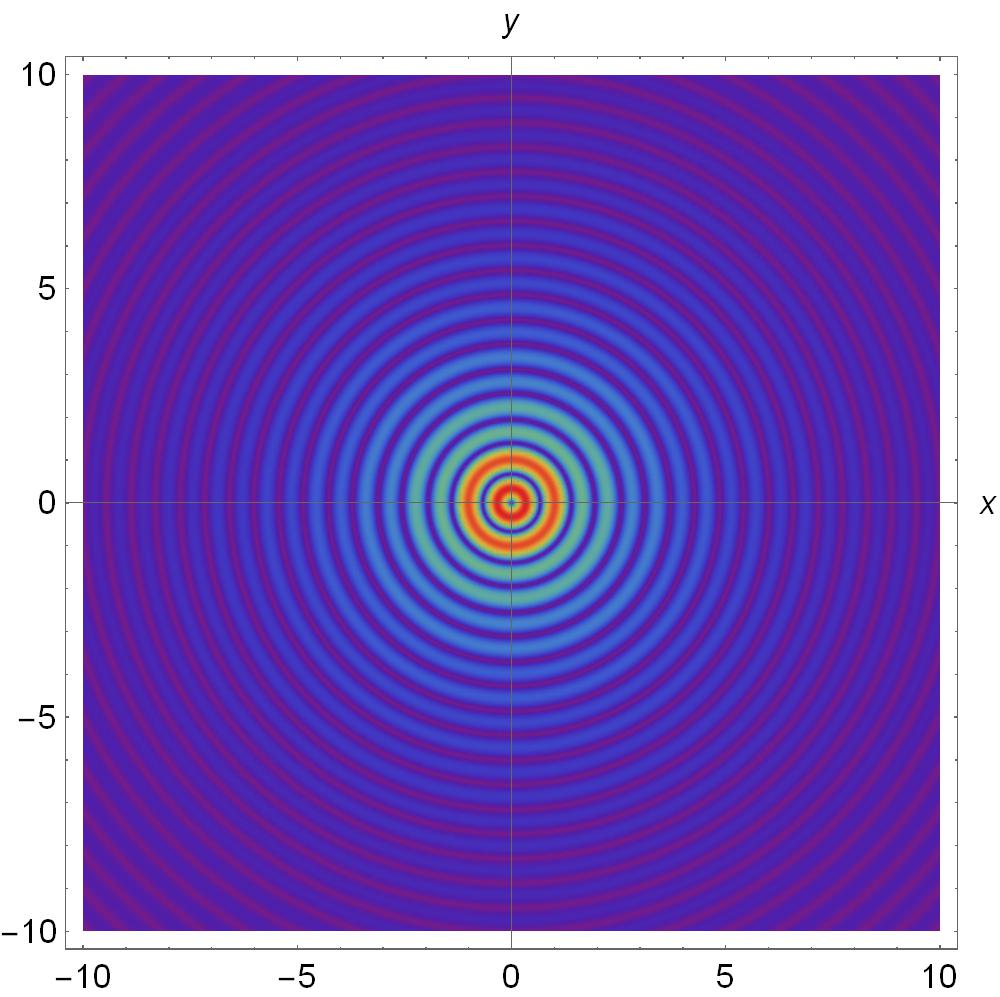} \\ 
	\caption{The transversal vector orbital angular momentum (left) and its intensity (right) for a Bessel beam with $m=1$ and $\sigma=i$.}
	\label{Figure3}
\end{figure}

\begin{figure}
	\centering
	\includegraphics[width=0.45\textwidth]{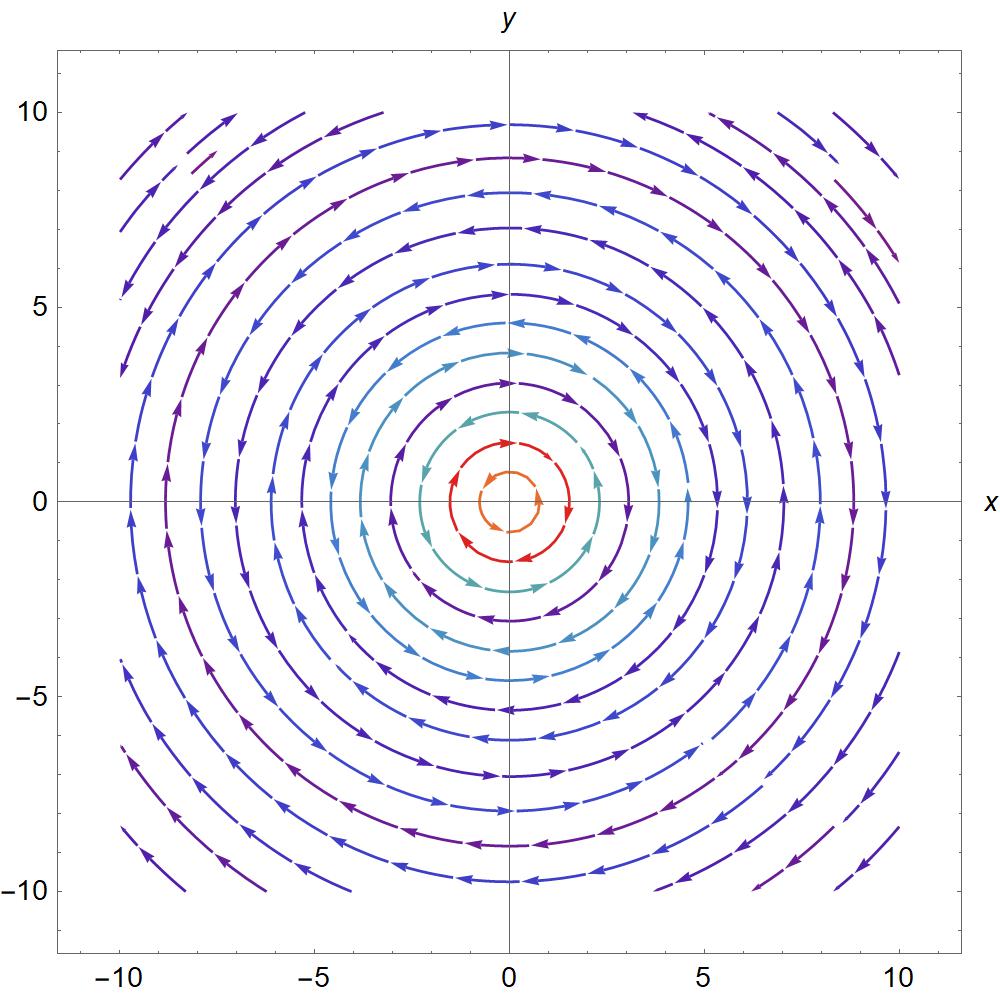} \qquad 
	\includegraphics[width=0.45\textwidth]{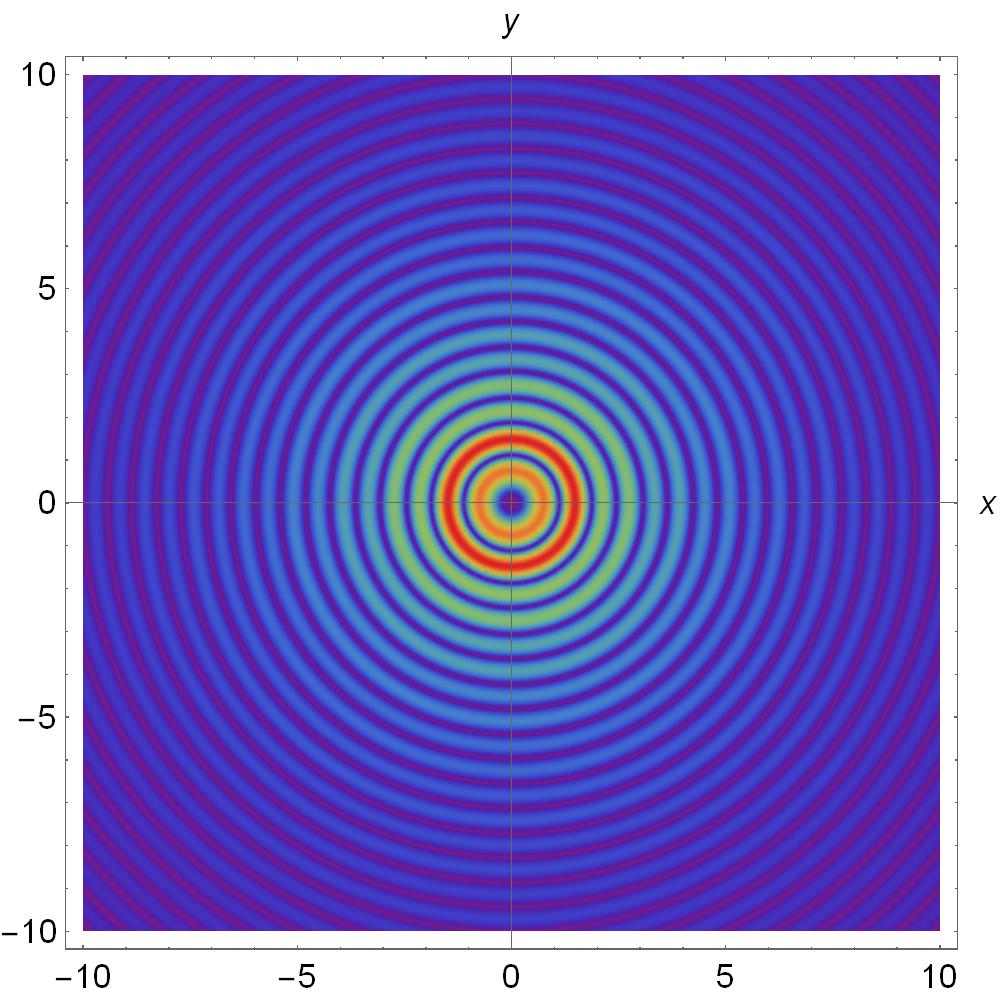} \\ 
	\caption{The transversal vector orbital angular momentum (left) and its intensity (right) for a Bessel beam with $m=2$ and $\sigma=-i$.}
	\label{Figure4}
\end{figure}

We can finally consider the paraxial approximation; i.e., the case when $k \approx k_z $. We substitute that approximation  into equation \eqref{Jtetm}, in which  the second term vanishes. Without loss of generality, we can  rewrite the complex constants  defined above as $c_{TE}\equiv i \alpha$ and $c_{TE}\equiv \beta$. Then, it is straightforward to obtain
\begin{align}\label{Jtetm}
\left\langle \vec{j}^{\,\,\text{TE/TM}}  \right\rangle  = & \frac{c \varepsilon }{2 \omega ^2}   \mathbf{Re} \Biggl\{ i \left( k^2 c_{\text{TE}} c^*_{\text{TM}}+ c^*_{\text{TE}} c_{\text{TM}} k_z^2 \right)\left(     \nabla_T \varphi^*\times \nabla_T \varphi  \right)    \cdot (\vec{r} \times  \hat{e}_3) 
+  c_{\text{TE}}^*\,c_{\text{TM}} k_z k_{T}^2   \,
\vec{r}\times [\hat{e}_3 \times \nabla_T   \left(  \varphi^* \varphi   \right)]  \Biggr\}
\end{align}
and
\begin{align}\label{ec:Jtapp}
\left\langle \vec{j}^{\,\,\text{TE/TM}}  \right\rangle =  \frac{ \varepsilon  }{2 c }   \mathbf{Re} \Biggl\{ \left(  \alpha^* \beta - \alpha \beta^*  \right)
\left(\nabla _{T}\varphi \cdot \nabla _{T}^{\perp} \varphi ^*\right)   (\hat{e}_1 u_2-\hat{e}_2 u_1)   \Biggr\}.
\end{align}
This expression clearly shows the vectorial transversal structure propagating due to mixed modes for any invariant beam. It is worth noticing the presence of  the term $\sigma \equiv \left(  \alpha^* \beta - \alpha \beta^*  \right)$,  which is due to the fact that any invariant beam can be generated as a superposition of plane waves \cite{NietoV}. This parameter $\sigma$ has been  physically linked  with polarization states and it has been usually reported in paraxial  and nonparaxial  approximations \cite{Alison,Padgett}. After substituting  a Bessel beam expressed by \eqref{BB} into \eqref{ec:Jtapp} and changing into cylindrical coordinates $x=r\cos\theta$ and $y=r\sin\theta$, we obtain 
\begin{align}\label{ec:IJtetmBB}
\left\langle \vec{j}^{\,\,\text{TE/TM}}  \right\rangle =   m\frac{ \varepsilon  }{2 c}
i \sigma k_T  J_{m }(k_T r)(J_{m +1}(k_T r)-J_{m -1}(k_T r)) (\hat{e}_r \sin\theta -\hat{e}_\theta \cos\theta).
\end{align}
This expression reveals that the orbital angular momentum contribution of the mixed modes is proportional to the topological azimuthal $m$ factor and depends linearly of the transversal wave vector. Taking  $\sigma=0$ means linear polarization and the lack of possible interference modes; otherwise, for $\sigma= \pm i$, we have circular polarization with  the existence of mixed modes. \\
Fig. \eqref{Figure3} shows the transverse amplitude vector with the mixed orbital angular momentum modes given by the equation  \eqref{ec:IJtetmBB}, for different  values of $m$, with $\sigma=i$. 
\noindent In Fig. \eqref{Figure4}, we change the azimuthal order to $m=2$ and the polarization is $\sigma=-i$. In both cases the intensity exhibits an azimuthally asymmetric shape which becomes circularly symmetrical.
\noindent Lastly, equation \eqref{ec:IJtetmBB} can be written using the following Bessel function identity $J_{m-1}(z)-J_{m+1}(z)= 2J_m^{\,'}(z)$ \cite{Abramowitz}, to obtain
\begin{align}
\left\langle \vec{j}^{\,\,\text{TE/TM}}  \right\rangle =&
-i \frac{\varepsilon}{c}\sigma 
m k_T J_{m }(k_T r)J_{m}^{\,'}(k_T r)
(\hat{e}_r \sin\theta -\hat{e}_\theta \cos\theta),
\nonumber \\ 
=&-i \frac{\varepsilon}{2c} \sigma 
m \frac{dJ^2_{m }(k_T r)}{dr}
(\hat{e}_r \sin\theta -\hat{e}_\theta \cos\theta).
\end{align}
It is possible to  identify the input field intensity $I(r)$ as $I(r)= \vert \vec{E}\vert^2= J^2_{m }(k_T r)$, and using $c=\omega/k $, we get
\begin{align}
 \label{Eq:vecJtetm}
\left\langle \vec{j}^{\,\,\text{TE/TM}}  \right\rangle =-i  \frac{ \varepsilon \sigma  }{2} m \frac{k}{\omega}
\frac{dI(r)}{dr} 
(\hat{e}_r \sin\theta -\hat{e}_\theta \cos\theta).
\end{align}
Thus, the transversal mixed modes orbital angular momentum can be calculated as the radial derivative of the input field intensity with a well define topological charge. Notice also that this expression contains the optical parameter, and it is proportional to  the topological charge $m$. It is worth mentioning that a similar term was reported theoretically in \cite{Zenkova}  and tested  experimentally in \cite{Angelsky}, to explain  the mechanical action of the spin part of the internal energy flow. There, the authors also showed the possibility of controllable motion of suspended particles by changing the polarization of the input field.

\section{Conclusions}
We have investigated  the transversal and longitudinal propagation of orbital angular momentum for invariant beams using a single scalar potential.  We have proved that the invariant beams satisfy Maxwell equations and possess well-defined orbital angular momentum \cite{Brandao}. We have shown that the superposition of non-paraxial vectorial beams with axial symmetry can provide a well defined orbital angular momentum. These results exhibit how the modes superposition affects the angular momentum flux density and causes reverse propagation in the case of the fractional Bessel beams \cite{FGMitri1,FGMitri2}. In \cite{Artur}, the authors have studied the importance to handle the amplitude, phase and polarization  in order to  design structured fields to study the spin-orbit interactions in Bessel beams. Adopting the single scalar potential approach, presented here, may be useful to find interesting applications, as the electromagnetic spin and canonical momentum for paraxial and nonparaxial beams recently reported in \cite{KYBliokh1,KYBliokh2,KYBliokh3}.

\end{document}